\documentclass[11pt]{article}
\usepackage{graphicx}
\usepackage{amsmath}
\usepackage{amssymb}
\newcommand{\BABARPubYear}    {03}

\newcommand{\BABARConfNumber} {25}
\newcommand{\SLACPubNumber} {10104}
\newcommand{\LANLNumber} {****}

\input pubboard/babarsym
\def\bKForg {\rm {$B\ra K_2^*(1430) \gamma $}}
\def\bKForgn {\rm {$B^0\ra K_2^{*}(1430)^0 \gamma $}}
\def\bKForgc {\rm {$B^{+}\ra K_2^{*}(1430)^+ \gamma $}}

\newcommand{\bKThrg}{\rm $B\ra K^*(1410) \gamma $}

\def\cth	{\mbox{$|\cos{\theta_H}|\ $}}

\def\etal{{\it et al.}}
\long\def\inst#1{\par\nobreak\kern 4pt\nobreak
    {\it #1}\par\vskip 10pt plus 3pt minus 3pt}

\begin{document}
{\pagestyle{empty}

\begin{flushright}
\babar-CONF-\BABARPubYear/\BABARConfNumber \\
SLAC-PUB-\SLACPubNumber \\
hep-ex/\LANLNumber \\
August 2003 \\
\end{flushright}

\par\vskip 5cm

\begin{center}
\Large \bf {Measurement of the $B^{0}\ra K_2^{*}(1430)^0 \gamma \ $ and $B^{+}\ra K_2^{*}(1430)^+ \gamma \ $branching fractions  }
\end{center}
\bigskip

\begin{center}
\large The \babar\ Collaboration\\
\mbox{ }\\
\today
\end{center}
\bigskip \bigskip

\begin{center}
\large \bf Abstract
\end{center}
$\ \ \ $We have investigated the exclusive, radiative $B$-meson decay to the charmless meson $K_2^*(1430)$ in $88.5\times 10^6$ \BB  events accumulated between 1999 and 2002 with the \babar$\ $detector at the PEP-II storage ring. We present a preliminary measurement of the branching fractions \BR (\bKForgn) $= (1.22\pm0.25\pm0.11)\times10^{-5}$ and \BR (\bKForgc)$ = (1.44\pm0.40\pm0.13)\times10^{-5}$, where the first error is statistical and the second systematic.
\vfill
\begin{center}
Contributed to the 
XXI$^{\rm st}$ International Symposium on Lepton and Photon Interactions at High~Energies, 8/11 --- 8/16/2003, Fermilab, Illinois USA
\end{center}

\vspace{1.0cm}
\begin{center}
{\em Stanford Linear Accelerator Center, Stanford University, 
Stanford, CA 94309} \\ \vspace{0.1cm}\hrule\vspace{0.1cm}
Work supported in part by Department of Energy contract DE-AC03-76SF00515.
\end{center}

\newpage
}

\begin{center}
\small

The \babar\ Collaboration,
\bigskip

%
B.~Aubert,
R.~Barate,
D.~Boutigny,
J.-M.~Gaillard,
A.~Hicheur,
Y.~Karyotakis,
J.~P.~Lees,
P.~Robbe,
V.~Tisserand,
A.~Zghiche
\inst{Laboratoire de Physique des Particules, F-74941 Annecy-le-Vieux, France }
% ...

\end{center}\newpage

\section{INTRODUCTION}
\label{sec:Introduction}

\begin{figure}[htb]
\begin{center}   
\begin{tabular}{c} 
   \mbox{\includegraphics[width=5in]{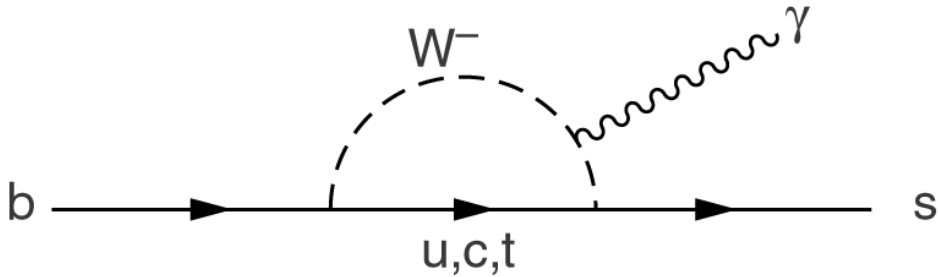}}
\end{tabular}
\end{center}    
\caption[fig:penguin]{$b\rightarrow s\gamma$ ``penguin'' diagram.
\label{fig:penguin}}
\end{figure}

$\ \ \ $In the Standard Model, flavor-changing neutral currents (FCNC) are forbidden at the tree level. For example, there is no direct coupling between the $b$ quark and the $s$ or $d$ quarks. Effective FCNC are induced by loop (or ``penguin'') diagrams, where a quark emits and re-absorbs a $W$ thus changing flavor twice, such as the $b\ra s\gamma$ transition depicted in Fig.~\ref{fig:penguin}.
    
With large samples of $B$-mesons and increasingly powerful background-suppression techniques, experimenters have succeeded in measuring Cabibbo-Kobayashi-Mashawa favored penguins. Theory has also made good progress in computing the decay rates; in particular, recent completion of the NLO correction~\cite{NLOtheo} to the inclusive $b\rightarrow s \gamma$ decay rate has sparked experimental efforts to further improve the measurement. Comparisons between theoretical and experimental rates place strong constraints on physics beyond the Standard Model~\cite{Model}. 

The discovery of $B\rightarrow K^*(892)\gamma$ by the CLEO collaboration~\cite{CLEOresult} verified the existence of penguins. The same publication also reported the evidence for \bKForg, later confirmed by the BELLE collaboration~\cite{BELLEr}. Measurements of the inclusive decay rate have been made by several experiments~\cite{pseudo}. Detailed knowledge about the decays to resonant modes with masses higher than $K^*(892)$, such as $B \ra K_2^{*}(1430) \gamma$ decay, will provide a better understanding of the inclusive $b\ra s\gamma$ branching fraction in terms of the sum over exclusive modes.

\section{THE \babar\ DETECTOR AND DATASET}
\label{sec:babar}

$\ \ \ $This study is based on 81.4 \invfb of data collected at the \FourS resonance (``on-resonance'') with the \babar\ detector at the \pep2\ asymmetric \ep(3.1\gev)-\en(9.0\gev) storage ring, corresponding to $88.5\times 10^6$ \BB pairs. We have also collected a data sample of $ 9.6$\invfb at 40\mev below the \FourS energy (``off-resonance''). The number of \BB meson pairs is determined from the ratio of the number of hadronic events to muon pairs in on- and off-resonance data. We assume that the \FourS decays equally to neutral and charged $B$-meson pairs. 

The \babar\ detector is described elsewhere~\cite{ref:babar}. The analysis described below makes use of charged track, $K_S$ and $\pi^0$ reconstruction, along with the charged particle identification. Charged particle trajectories are measured by a 5-layer double-sided silicon vertex tracker (SVT) and a 40-layer drift chamber, which also provide ionization measurements ($dE/dx$) used for particle identification. For charged tracks with momentum $p>1$\gevc, the measured transverse momentum with respect to the beam axis ($p_T$) has the resolution 
\begin{equation}
\frac{\sigma_{p_T}}{p_T} = 0.13\% p_T + 0.45\% ,
\end{equation}
where $p_T$ is measured in\gevc. 

Photons and electrons are measured in the barrel and forward end-cap electromagnetic calorimeter, consisting of 6580 Thallium-doped CsI crystals. The electromagnetic calorimeter resolution, $\sigma_E$, can be expressed as
\begin{equation}
\frac{\sigma_E}{E} = \frac{2.3\%}{E^{\frac{1}{4}}} { \oplus} 1.9\% ,
\end{equation}
where the energy $E$ is measured in\gev.

Charged particle identification is provided by the energy loss ($dE/dx$) in the tracking devices and by an internally reflecting ring-imaging Cherenkov detector (DIRC). The DIRC transports the Cherenkov light to a water-filled expansion volume equipped with approximately 11,000 photomultiplier tubes. A $K/\pi$ separation better than four standard deviations is achieved for momenta below 3\gevc. 

We use Monte Carlo simulations of the \babar$\ $ detector based on GEANT 4~\cite{Geant} to optimize our selection criteria and to determine signal efficiencies. These simulations take into account the varying detector conditions and beam backgrounds during the data-taking period.

\section{B CANDIDATE RECONSTRUCTION}
\label{sec:selection}

$\ \ \ $The $K_2^*(1430)$ is reconstructed from $K^+$, $K^0$, $\pi^+$ and $\pi^0$ candidates through the three modes $K_2^{*0}(1430)\ra K^+\pi^-$ and $K_2^{*+}(1430)\ra K^+\pi^0, K^0\pi^+$. $K^0$ mesons are only reconstructed from the decay $K^0_S\ra \pi^+\pi^-$. In this paper the charge conjugate decays are implied unless otherwise stated.

A photon candidate is defined as a localized energy maximum within the calorimeter acceptance $-0.74<\cos\theta<0.93$, where $\theta$ is the polar angle to the detector axis. It must be isolated by 25\cm from any other neutral candidate or track and have a lateral energy profile consistent with a photon shower. To suppress photons from $\pi^0 (\eta)$ decays, we veto any photon that combines with another photon of energy greater than $50\ (250)$ \mev to form a $\gamma\gamma$ invariant mass in the range $115 (508)\ < M_{\gamma \gamma } < 155\ (588)$ \mevcc. 

The $\pi^0$ candidates are reconstructed from pairs of photons, which have energy above 50 \mev and opening angle less than 36 degrees; the invariant mass of the two photons is required to be in the range $115 < M_{\gamma\gamma} < 150$\mevcc; the $\pi^0$ momentum is recalculated with a $\pi^0$ mass constraint to improve the energy resolution.

The $K^\pm$ and $\pi^\pm$ track candidates are required to be well reconstructed in the drift chamber and to originate from a vertex consistent with the \epem interaction point (IP). 
A track is identified as a kaon if it is projected to pass through the fiducial volume of the DIRC, and the cone of Cherenkov light is consistent in time and angle with a kaon of the measured track momentum. A charged pion is identified as a track that does not satisfy the criteria for a kaon or an electron.

The $K^0_S$ candidates are reconstructed from two oppositely charged tracks coming from a common vertex displaced from the IP by at least 0.2\cm. Each candidate is required to lie in a direction consistent with the $K^0_S$ momentum, while having an invariant mass within a $489<M_{\pi^+\pi^-}<507$\mevcc window around the nominal $K^0_S$ mass. 

The $K_2^*(1430)$ candidate is required to have a $K\pi$ invariant mass within 120 and 110\mevcc of the known $K_2^{*0}$ and $K_2^{*+}$ mass~\cite{ref:pdg2002}. For the $K^+\pi^-$ mode, we require that the two tracks originate from a common vertex with a probability of at least 0.001. 

To remove Bhabha, radiative-Bhabha and tau events, we require that the second-order Fox-Wolfram moment~\cite{FoxWolf} of the event is less than 0.9.

The $B$ candidates are reconstructed by  combining one  $K_2^*(1430)$ and one $\gamma$ candidate with energy $1.80 < E^*_{\gamma} < 2.75 $\gev in the rest frame of the \FourS. In this frame, the energy of the $B$-meson is given by the beam energy, $\sqrt{s}/2$, which is known much more precisely than the energy of the $B$ candidate. Therefore, to isolate the $B$-meson signal, we use two kinematic variables: the difference between the reconstructed energy of the $B$ candidate and the beam energy in the center-of-mass frame (\DeltaE), and the beam energy substituted mass (\mes), defined as
\[
m_{ES}^{\rm raw}=\sqrt{E_{\rm beam}^{2}-p_B^{*2}},
\]
where \[
  E_{\rm beam}=\sqrt{s}/2,
\]
\[ \vec{p}_{B}^*=\vec{p}_{K^*}^*+\vec{p}_{\gamma}^*. \]

For the modes containing a single photon candidate, namely $K^+\pi^-$ and $K^0_S\pi^+$, we adopt a technique from the CLEO analysis~\cite{CLEOresult}, which rescales the measured photon energy $E^*_{\gamma}$ with a factor $\kappa$, determined for each event, such that $E^*_{K^*} + \kappa E^*_{\gamma} - E^*_{beam}=0$; this improves the \mes resolution from 3.0 to 2.7\mevcc. If we find multiple candidates with $|\DeltaE|<0.3$\gev and \mes$>5.2$\gevcc in the same event, which happens in $3.1\%$, $6.3\%$ and $4.9\%$ of the events for the $K^+\pi^-$, $K_S^0\pi^+$ and $K^+\pi^0$ modes, respectively, we take the candidate that has the invariant mass closest to the $K_2^*(1430)$ mass~\cite{ref:pdg2002}.
 
The background has two components, one of which is continuum \qqbar production, where $q$ can be a $u$, $d$, $s$ or $c$ quark, with the high-energy photon originating from initial-state radiation or from $\pi^0$ and $\eta$ decays. (This is known as the ``non peaking'' background in the \mes and \DeltaE distributions.)

The second background contribution is from other $B\ra X_s\gamma$ modes (dominantly \bKThrg) and non resonant $B\ra K\pi\gamma$ decays. We label these the ``peaking'' background, since these decays have \mes and \DeltaE distributions similar to the signal.  

In order to distinguish the \bKForg$\ $signal from \bKThrg$\ $and non resonant decays, we examine the helicity angle distributions. The helicity angle $\theta_H$ is defined as the angle of the $K^+$ or $\KS$ in the rest frame of the $K_2^*(1430)$ with respect to the flight direction of the $K_2^*(1430)$. All three modes have different helicity angle distributions: ${\sin}^2\theta_{H}{\cos}^2\theta_{H}$ for $K_2^*(1430)$, ${\sin}^2\theta_{H}$ for $K^*(1410)$ and primarily ${\sin}^2\theta_{H}$ for non resonant decays assuming $J=1$ for the spin of the $K\pi$ system. The non resonant decays may have higher angular momentum contributions but the lowest possible angular momentum state is dominant, so the helicity angle distribution for non resonant decay is attributed to be the same as that of \bKThrg$\ $ decay. The systematic uncertainty related with this modeling is studied and included in the measured branching fraction uncertainty. 

\section{BACKGROUND SUPPRESSION}

\begin{figure}[htb]
\begin{center}   
\begin{tabular}{cc} 
\mbox{\includegraphics[width=5in]{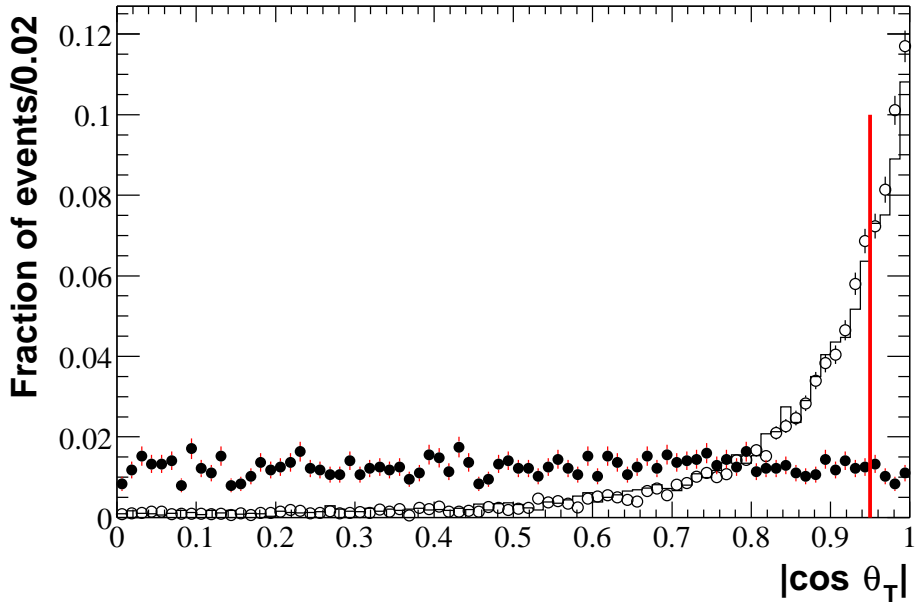}}
\end{tabular}
\end{center}    
\caption[fig:cosT]{ Thrust angle distribution: Filled circles correspond to \bKForg$\ $signal MC simulation, the histogram corresponds to off-resonance data, and the open circles correspond to continuum background MC simulation. The vertical line indicates the cut value.
\label{fig:cosT}}
\end{figure}

\begin{figure}[htb]
\begin{center}   
\begin{tabular}{cc} 
\mbox{\includegraphics[width=5in]{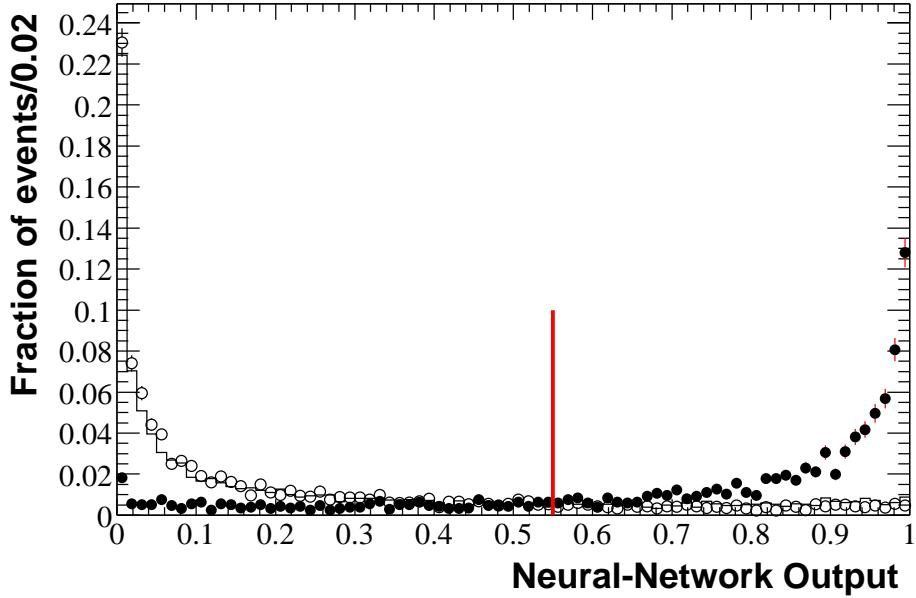}}
\end{tabular}
\end{center}    
\caption[fig:NNoutput]{ Neural network output distribution: Filled circles correspond to \bKForg$\ $signal MC simulation, the histogram corresponds to off-resonance data, and the open circles correspond to continuum background MC simulation. The vertical line indicates the cut value.
\label{fig:NNoutput}}
\end{figure}

$\ \ \ $We have exploited the difference in the event topology between signal and continuum background to reduce the continuum contribution, as well as the combinatorial \BB background. The distribution of the thrust angle, defined as the angle between the direction of the photon candidate and the thrust axis of the rest of the event (the particles not used in the reconstruction of the B candidate) in the center-of-mass frame, is shown in Fig.~\ref{fig:cosT}. We train a neural network with a combination of the thrust angle, the angle of the $B$-meson candidate's direction with respect to the beam axis, the center-of-mass momentum vectors for tracks and photons not associated with the $B$ candidate (binned in $10^{\circ}$ intervals ranging from parallel to anti parallel relative to the photon momentum), sphericity and the ratio of second-to-zeroth order Fox-Wolfram moments in the photon recoil system. The neural network improves background suppression significantly. The distribution of the neural network output is shown in Fig.~\ref{fig:NNoutput} for MC signal, MC continuum background and off-resonance data. The neural network software used for this analysis is based on the Stuttgart Neural Network Simulator~\cite{SNNSNET}.

The cuts on thrust angle ($\theta_{\rm T}$) and neural network output (NNO) have been optimized using an iterative method to minimize correlations. The final cuts are $|\cos\theta_{\rm T}| < 0.95$ and NNO$ > 0.55$, shown in Figs.~\ref{fig:cosT} and \ref{fig:NNoutput}. Because the variables used for the neural network training are mostly calculated from the rest-of-the-event information, we use a sample of fully reconstructed $B\ra D\pi^-$ candidates in data, as well as a sample of simulated $B\ra D\pi^-$ events as control samples. The bachelor pion in the $B\ra D\pi^-$ decay is treated like the photon in \bKForg$\ $decay for the calculations of the event variables; the difference in the efficiency of the cut on the neural network output between different samples is taken as the systematic uncertainty related to this cut.

\section{ANALYSIS METHOD}
\label{sec:Analysis}
$\ \ \ $An unbinned maximum-likelihood technique is used to fit simultaneously the \mes, \DeltaE and \cth distributions. The fit is performed independently for each of the decay modes considered here.

The signal \mes and \DeltaE distributions are well described by an asymmetric resolution function (``Crystal-Ball" function)~\cite{Crystal} $S$, having an approximately Gaussian core plus a long tail due to the energy leakage from the calorimeter for the photon candidates:
\begin{equation}
S \propto
\begin{cases}
 \exp(-(m-m_0)^2/(2{\sigma}^2))& \text{for $m>m_0-\alpha \sigma$} ,\\
\frac{{(n/\alpha )}^n exp(-{\alpha}^2/2)}{({(m_0-m)/\sigma + n/\alpha -\alpha)}^n}& \text{for $m\leqslant m_0 - \alpha \sigma$}.
\end{cases}
\end{equation}
The continuum background is parameterized empirically by a threshold function~\cite{BGpdf} for \mes and a linear function for \DeltaE.

The $\cos{\theta_H}$ distribution of the signal has been parameterized with ${\sin}^2{\theta_H}{\cos}^2{\theta_H}-\lambda({\cos}^4{\theta_H}-{\cos}^6{\theta_H})$, where $\lambda$ is a parameter determined from the Monte Carlo sample to account for the effect of the detection acceptance and efficiency. The \cth distribution of the ``non peaking'' background is parameterized by the combination of exponential and constant components. 
 
\begin{figure}[htb]
\begin{center}   
\begin{tabular}{c} 
   \mbox{\includegraphics[width=5.0in]{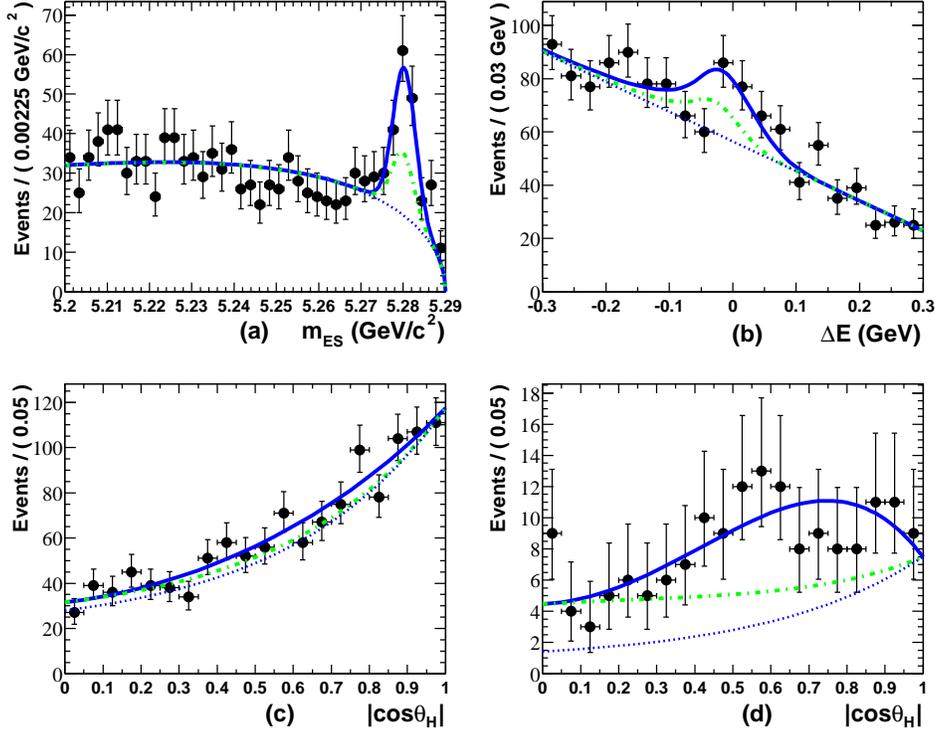}}
\end{tabular}
\end{center} 
\caption[fig:kpionr]{A fit to the (a) \mes, (b) \DeltaE and (c) \cth distributions for \bKForgn, $K_2^{*}(1430)^0\ra K^+\pi^-$ candidates in data, and (d) the \cth distribution in the signal region. The solid line shows a fit to the data. The peaking (dash-dot) and non peaking (dash) background contributions are also shown.
\label{fig:kpionr}}
\end{figure}

\begin{figure}[htb]
\begin{center}   
\begin{tabular}{c} 
   \mbox{\includegraphics[width=5.0in]{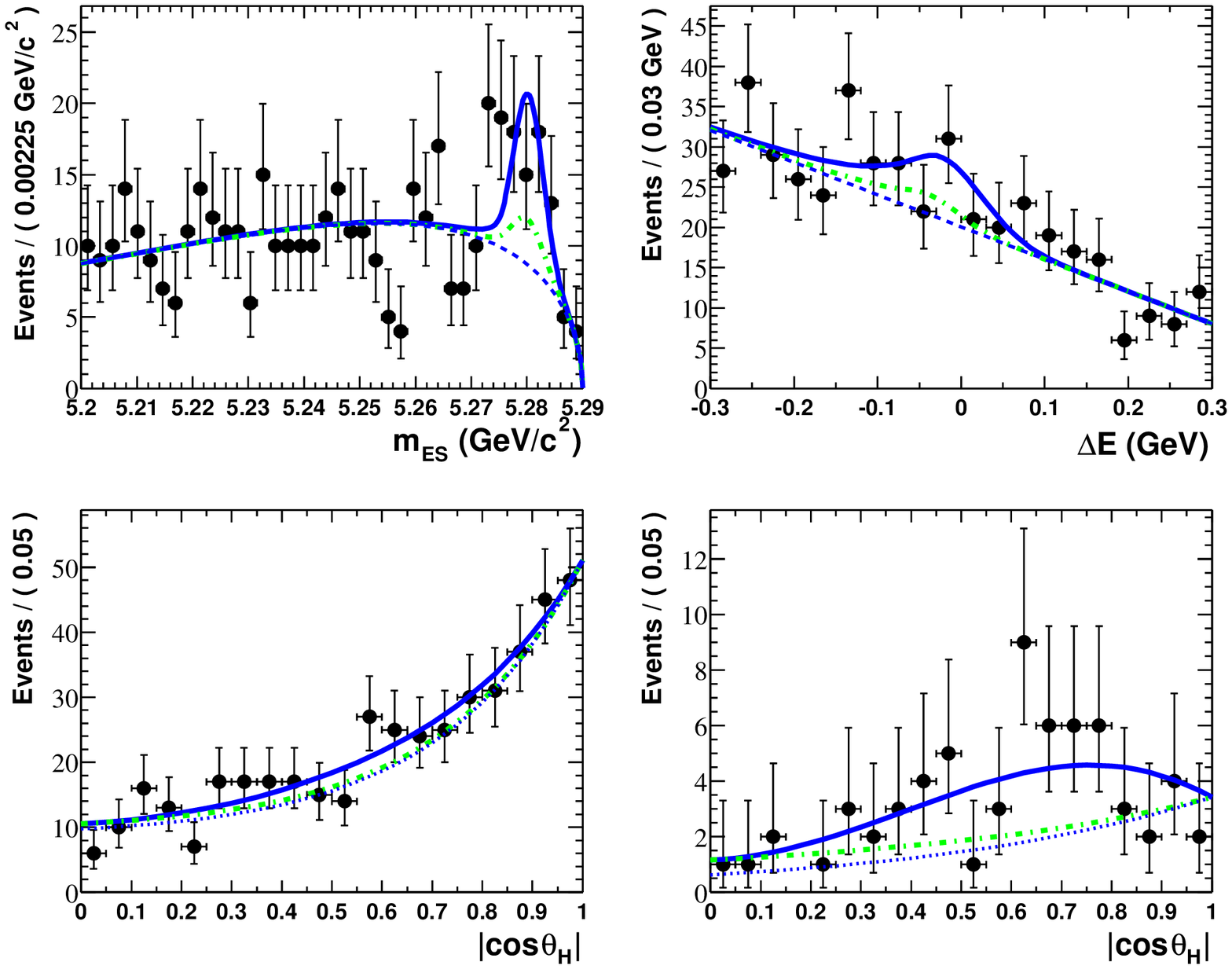}
	}
\end{tabular}
\end{center} 
\caption[fig:kspionr]{A fit to the (a) \mes, (b) \DeltaE and (c) \cth distributions for \bKForgc, $K_2^{*+}(1430)\ra K_S^0\pi^+$ candidates in data, and (d) the \cth distribution in the signal region. The solid line shows a fit to the data. The peaking (dash-dot) and non peaking (dash) background contributions are also shown.
\label{fig:kspionr}}
\end{figure}

\begin{figure}[htb]
\begin{center}   
\begin{tabular}{c} 
   \mbox{\includegraphics[width=5.0in]{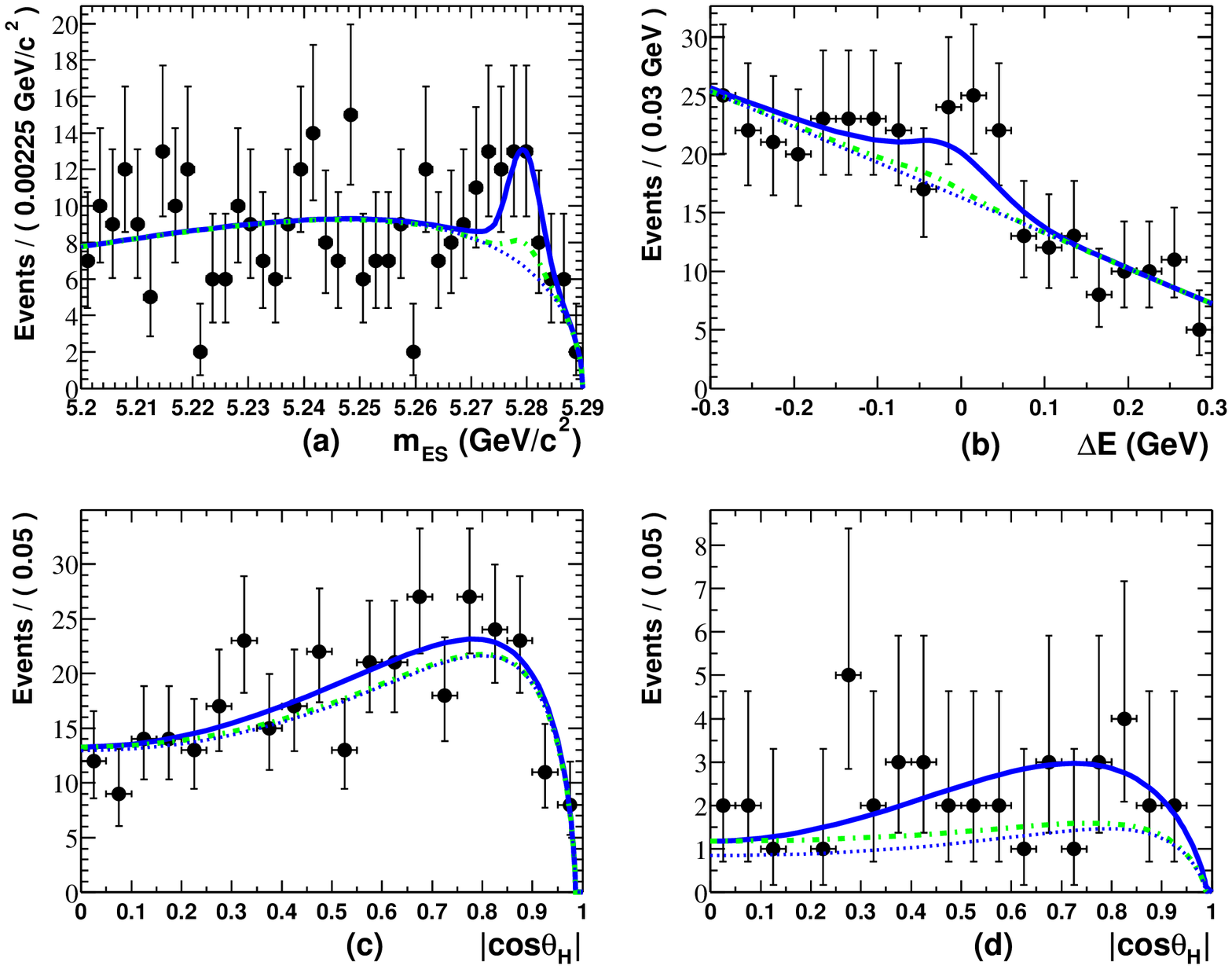}
	}
\end{tabular}
\end{center} 
\caption[fig:kpizonr]{A fit to the (a) \mes, (b) \DeltaE and (c) \cth distributions for \bKForgc,$\ K_2^{*+}(1430)\ra K^+\pi^0$ candidates in data, and (d) the \cth distribution in the signal region. The solid line shows a fit to the data. The peaking (dash-dot) and non peaking (dash) background contributions are also shown.
\label{fig:kpizonr}}
\end{figure}

Figures.~\ref{fig:kpionr}, \ref{fig:kspionr} and \ref{fig:kpizonr} show the \mes, \DeltaE and \cth distributions for the $K^+\pi^-$, $K^0_S\pi^+$ and $K^+\pi^0$ modes, respectively. The \cth distributions in the signal region (defined as $-0.15< \DeltaE <0.10$\gev, $5.272<\mes<5.288$\gevcc) are also shown. The signal as well as background yields are allowed to vary in the fit. All the non peaking background parameters are allowed to float. The signal and peaking-background helicity angle, Crystal-Ball width and shape parameters are constrained according to the MC expectations. The means of the signal \mes and \DeltaE functions are constrained to the MC expectations, calibrated using $B\ra K^*(892)\gamma$ candidates from MC simulation and data, while the peaking-background means are allowed to float due to its complex composition. The signal and peaking-background yields are given in Table~\ref{tab:yields}. The signal significance has been evaluated from the change in the likelihood when the fit is repeated with the signal yield set to zero.

\begin{figure}[htb]
\begin{center}   
\begin{tabular}{c} 
   \mbox{\includegraphics[width=3.0in]{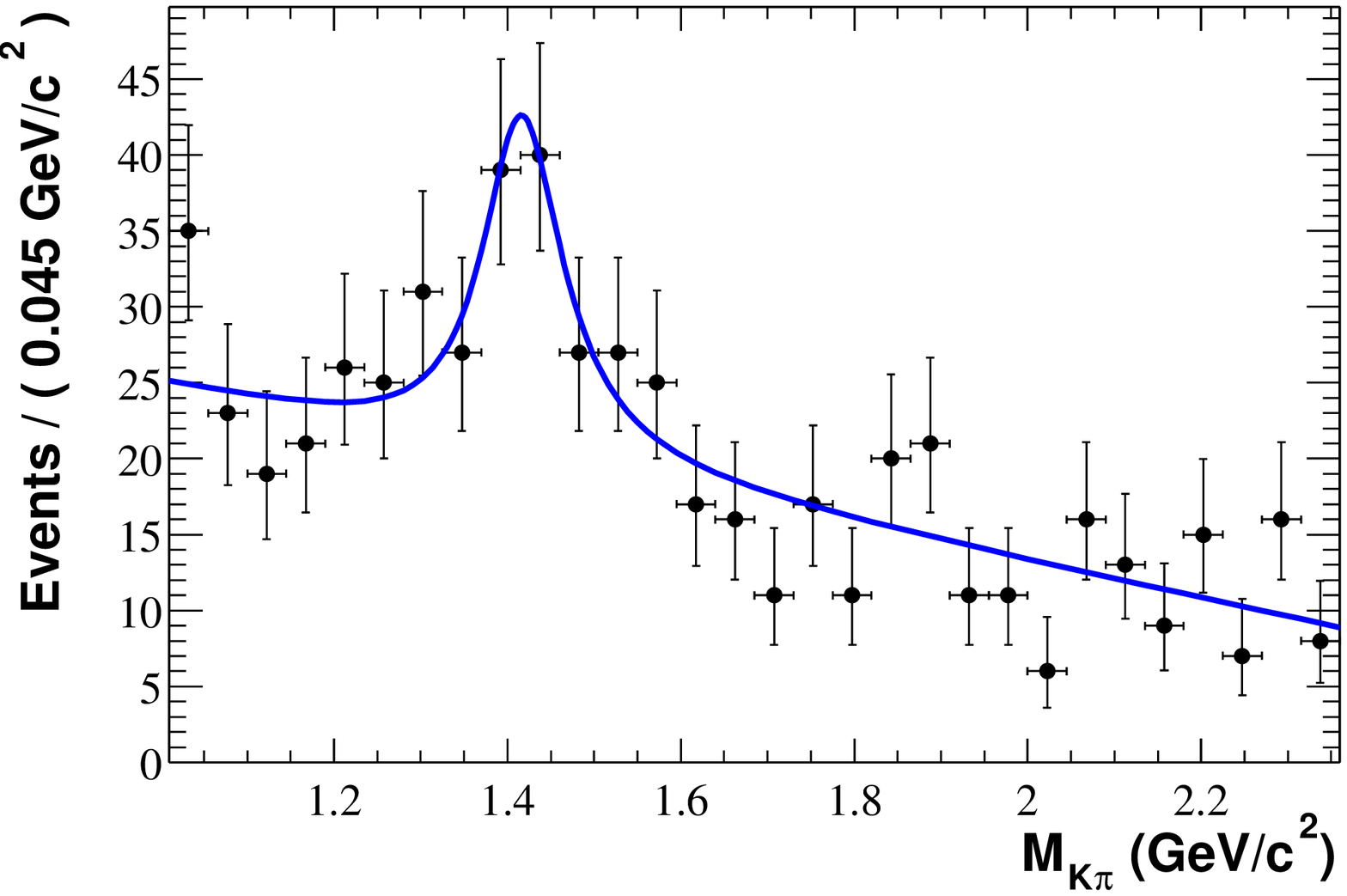}
	\includegraphics[width=3.0in]{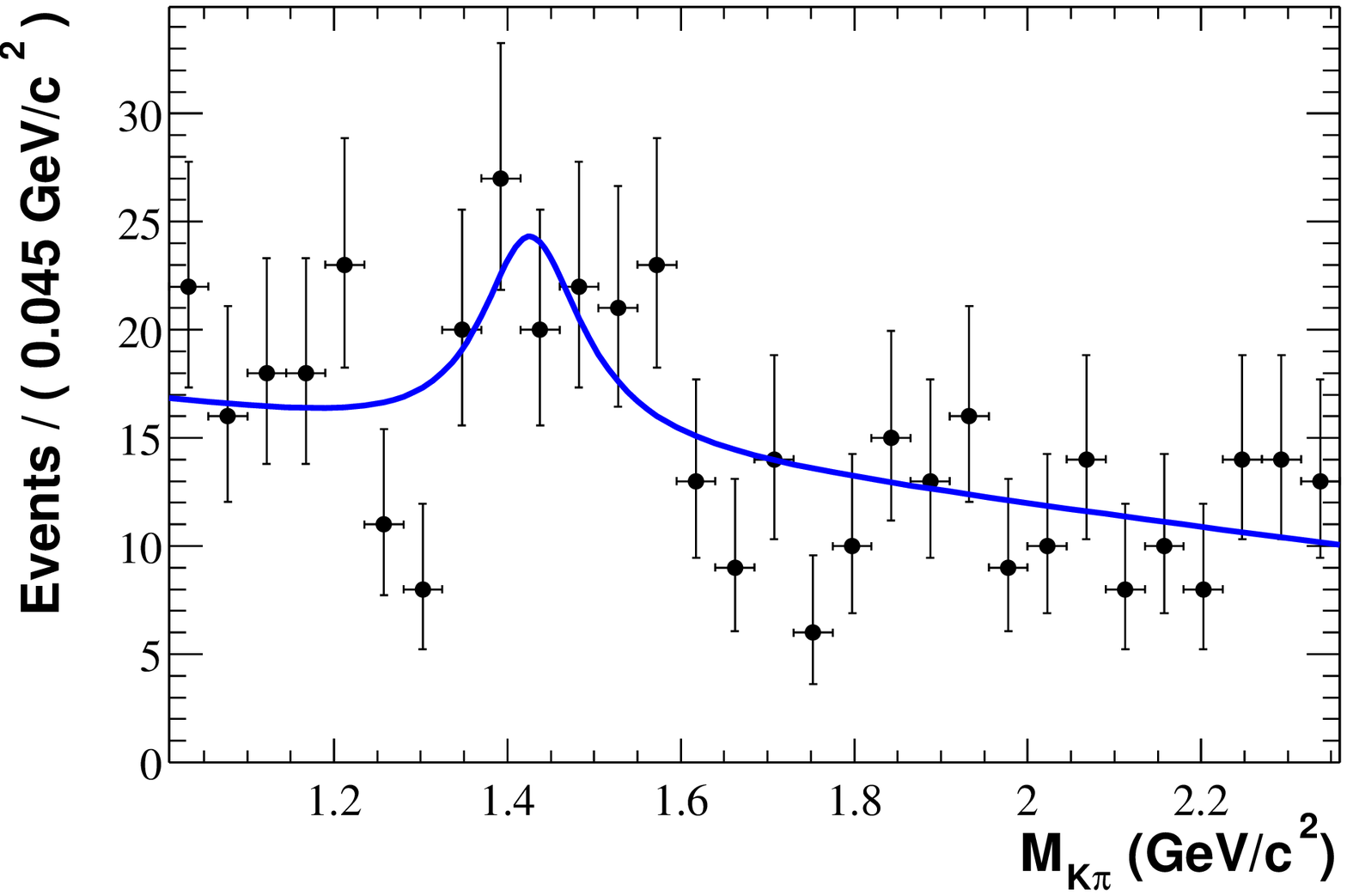}}
\end{tabular}
\end{center} 
\caption[fig:mksproject]{$K^+\pi^-$ (left) and $K^0_S\pi^+$ and $K^+\pi^0$ (right) invariant mass distributions for the signal region (see text for a definition of the signal region).
\label{fig:mksproject}}
\end{figure}

\begin{table}
\caption{The efficiency, fitted signal yield, peaking-background yield and measured branching fraction \BR (\bKForg) for each $K_2^{*}(1430)$ decay mode.}
\begin{center}
\begin{tabular}{|c|c|c|c|c|c| }      \hline
 Mode & Efficiency & Signal   & Peaking  & Signal & \BR (\bKForg)\\  
      & ($\%$)       &        &   background & significance  & ($10^{-5}$) \\ \hline
 $K^+\pi^-$  & 6.4  & $69.\pm14.$   & $47.\pm12.$ & 5.8 & $1.22\pm0.25\pm0.11$ \\  \hline
 $K^0_S\pi^+$ & 1.9  & $29.\pm10.$  & $11.\pm8.$ & 3.3  & $1.69\pm0.59\pm0.20$ \\ \hline 
 $K^+\pi^0$  & 1.9     &$20.\pm9.$ & $4.\pm7.$ & 2.6  & $1.23\pm0.55\pm0.12$ \\ \hline 
\end{tabular} 
\end{center}
\label{tab:yields}
\end{table}
Figure.~\ref{fig:mksproject} shows the $K\pi$ invariant mass distribution, which is fit with a relativistic Breit-Wigner function plus a first-order polynomial background, though we understand that it is not the most accurate description of the resonance due to possible interferences. There is a clear enhancement around 1.4\gevcc in the neutral mode and a slight enhancement in the charged modes.

\section{SYSTEMATIC STUDIES}
\label{sec:Systematics}
$\ \ \ $The total systematic error is the sum in quadrature of the components shown in Table~\ref{tab:syst}. The \DeltaE resolution is dominated by the photon energy resolution, which is determined from data using $\pi^0$ and $\eta$ meson decays with symmetric decay-photon energies. The deviation in the reconstructed $\eta$ mass from the nominal $\eta$ mass provides an estimate of the uncertainty in the measured single photon energy. The photon isolation and $\pi^0/\eta$ veto efficiency depends on the event multiplicity and the effect is estimated by ``embedding'' MC-generated photons into both an exclusively reconstructed $B$-meson data sample and a generic $B$-meson MC sample. The photon and $\pi^0$ efficiency uncertainties are determined from a comparison of the efficiencies in data and MC for $\epem\ra\tautau$ events. The tracking-efficiency uncertainty is estimated from a sample of tracks well measured in the SVT. We estimate the uncertainties in the $K^0_S$ efficiency by comparing the data and MC distributions of the momentum and flight distance. The efficiency for kaon identification in the DIRC is derived from a sample of the decays $D^{*+}\ra D^0\pi^+$, with $D^0\ra K^-\pi^+$. 

The uncertainties in the background suppression cuts (the thrust angle and neural network cuts) are obtained by comparing the cut efficiencies between the \bKForg$\ $MC, and $B\ra D\pi^+$ data and MC samples, and choosing the largest discrepancy. The invariant mass and width used for $K_2^*(1430)$ are from the PDG~\cite{ref:pdg2002}. The systematic error is obtained by varying the mean and width, within their errors, and using that which results in the largest change in efficiency.

We estimate the systematic error due to the fitting procedures as follows. For the shape parameters for \mes, \DeltaE and \cth distributions, we vary the parameters in the fit within their errors from the MC expectations. We also test the validity of the peaking-background \cth probability density function (PDF) by replacing it with different parameterized PDFs from MC samples. We double the largest deviation in these tests as the systematic error of the signal yield. There is also a small systematic error associated with the limited statistics of the signal MC sample.
\begin{table}[!htb]
\caption{Fractional systematic uncertainties ($\%$) in the measurement of \BR(\bKForg).}
\begin{center}
\begin{tabular}{|l|c|c|c|}  \hline
Uncertainty & \multicolumn{3}{c|}{ } \\ \cline{2-4} 
                          & $\Kpm \pimp$ & $\KS \pipm$ & $\Kpm \piz$ \\ \hline 

$B$-counting                               & 1.1      &1.1    & 1.1  \\
Photon detection efficiency              & 1.3       &1.3    &1.3  \\
Photon energy scale                      & 1.0       & 1.0   & 1.0 \\
Photon energy resolution                 & 2.5       & 2.5   & 2.5 \\
Photon isolation                      & 2.0       & 2.0   & 2.0 \\
$\pi^0/\eta$ veto                        & 1.0       & 1.0   & 1.0 \\
$K^+/\pi^+$ tracking                 & 1.6       & 0.8   & 0.8 \\
$K^0_S$ efficiency                       & ...       & 3.0   & ...\\
$\pi^0$ efficiency                       & ...       & ...   & 2.5 \\
Sub-mode branching fraction                    & 2.4      & 2.4   & 2.4 \\
$K_2^*(1430)$ mass/width                        & 1.6      & 1.0   & 1.1   \\
Signal PDF parameters                    & 3.9      & 5.8    & 6.3   \\
Background suppression                   & 5.0     & 6.0   & 6.0 \\
Peaking-background modeling             & 3.0      & 5.0   & 5.0   \\
MC statistics                            & 2.5      & 3.2   & 3.2  \\ \hline
Total                                    & 9.      & 12.   & 10.  \\ \hline
\end{tabular}
\end{center}
\label{tab:syst}
\end{table}

\section{CONCLUSIONS}
\label{sec:Physics}

$\ \ \ $We have presented a preliminary measurement of the branching fraction for \bKForgn$\ $of $(1.22\pm0.25\pm0.11)\times10^{-5}$ with a $5.8\sigma$ statistical significance, which is an improvement over previous experimental results, while still being in agreement. We observe a signal with a statistical significance of $4.1\sigma$ for \bKForgc$\ $ and measure the preliminary branching fraction to be $(1.44\pm0.40\pm0.13)\times10^{-5}$, by combining the results from $K_S^0\pi^+$ and $K\pi^0$ modes. Both results agree with the theoretical prediction from the relativistic form-factor model of Veseli and Olsson~\cite{Theo}.

\section{ACKNOWLEDGMENTS}
\label{sec:Acknowledgments}

$\ \ \ \ \ $
We are grateful for the 
extraordinary contributions of our \pep2\ colleagues in
achieving the excellent luminosity and machine conditions
that have made this work possible.
The success of this project also relies critically on the 
expertise and dedication of the computing organizations that 
support \babar.
The collaborating institutions wish to thank 
SLAC for its support and the kind hospitality extended to them. 
This work is supported by the
US Department of Energy
and National Science Foundation, the
Natural Sciences and Engineering Research Council (Canada),
Institute of High Energy Physics (China), the
Commissariat \`a l'Energie Atomique and
Institut National de Physique Nucl\'eaire et de Physique des Particules
(France), the
Bundesministerium f\"ur Bildung und Forschung and
Deutsche Forschungsgemeinschaft
(Germany), the
Istituto Nazionale di Fisica Nucleare (Italy),
the Foundation for Fundamental Research on Matter (The Netherlands),
the Research Council of Norway, the
Ministry of Science and Technology of the Russian Federation, and the
Particle Physics and Astronomy Research Council (United Kingdom). 
Individuals have received support from 
the A. P. Sloan Foundation, 
the Research Corporation,
and the Alexander von Humboldt Foundation.

\end{document}